\documentclass[12pt]{iopart}
\newcommand{\beq}{\begin{equation}}
\newcommand{\eeq}{\end{equation}}

\usepackage{iopams,graphicx}  
\expandafter\let\csname equation*\endcsname\relax
\expandafter\let\csname endequation*\endcsname\relax
\usepackage{amsmath}
\begin{document}
\title{Chiral propulsion by electromagnetic fields}
\author{S. Aif}
\address{Institute of High Technologies, Taras Shevchenko National University of Kyiv, \\ 03022 Kyiv, Ukraine}
\ead{sergiyayf@gmail.com}
\author{I.A. Kuk}
\address{Faculty of Avionics, Energy and Infocommunications, \\ Ufa State Aviation Technical University, Ufa 450000, Russia}
\ead{ilyha13@gmail.com}
\author{D.E. Kharzeev}
\address{Department of Physics and Astronomy, Stony Brook University, \\ New York 11794, USA}
\vskip0.3cm
\address{Center for Chiral Biophotonics, School of Biomedicine, \\ Far Eastern Federal University, Vladivostok 690000, Russia}
\ead{dmitri.kharzeev@stonybrook.edu}
\vspace{10pt}
\begin{indented}
\item[]17 April 2018
\end{indented}

\begin{abstract}
We consider the propulsion of micron-scale chiral objects by electromagnetic fields in fluids -- a problem with broad applications in microfluidics, pharmaceutics, and biomedicine. Because of the small size of the moving objects, the propulsion can be described by the Stokes equation possessing the time-reversal invariance. We propose a method of evaluating the propulsion velocity based on the Green's function of the Stokes equation. As an illustration, we first use it to provide a simple derivation of the classic Stokes law for a sphere moving in a viscous fluid. We then use this method for describing the
propulsion of helical bodies, evaluate the propulsion velocity, and find that it does not depend on the viscosity of the fluid as long as the Reynolds number remains small. As an application, we describe 
 recent experimental results on the propulsion of nano-propellers by electromagnetic fields in water, with a good agreement with the data. We also discuss applications to optofluidic chiral sorting of molecules by rotating electromagnetic fields and circularly polarized light.
\end{abstract}

\section{Introduction}

Chirality -- the lack of symmetry between left and right -- is a ubiquitous concept in physics, chemistry, and biology. Chirality is closely related to motion -- for example, rotation of a rigid chiral screw in the viscous fluid creates a translational motion, the effect utilized by Archimedes in the 3rd century BC. Long before Archimedes, Nature realized how to use this effect to propel bacteria and other microorganisms. 
\vskip0.3cm

It is instructive to analyze the motion of Archimedes' screw from the point of view of discrete symmetries. The translational velocity ${\vec V}$ of the screw is a vector, i.e. a quantity that is odd under parity transformation ${\cal P}$ that interchanges left and right. On the other hand, the angular momentum of the screw ${\vec \Omega}$ is a pseudo-vector that is even under ${\cal P}$. Therefore, the translational motion of a rotating screw described by the relation
\beq\label{propel}
{\vec V} = \alpha \ {\vec \Omega}
\eeq
necessarily requires the breaking of parity invariance, i.e. the rotating object has to possess chirality. In other words, the quantity $\alpha$ has to be ${\cal P}$-odd. 
\vskip0.3cm

One can also observe that both the velocity ${\vec V}$ and the angular momentum 
${\vec \Omega}$ are odd under time-reversal transformation ${\cal T}$ -- if the film is rolled backwards, both the velocities and the angular momenta change signs. This means that the quantity $\alpha$ in (\ref{propel}) is even under ${\cal T}$, and the translational motion of the screw does not have to be accompanied by the breaking of time reversal invariance. This observation is not trivial, since the propulsion of a rotating screw in the fluid requires a non-zero viscosity, and viscosity describes  dissipative processes. Dissipation leads to the production of entropy, and thus to the emergence of the arrow of time that breaks the time reversal invariance.  
Nevertheless, the symmetry properties of (\ref{propel}) allow the propulsion of chiral objects without dissipation. In particular, the ${\cal T}$-even coefficient $\alpha$ in (\ref{propel}) cannot be (inversely) proportional to the ${\cal T}$-odd viscosity. 
This property of chiral propulsion is in sharp contrast to the Stokes law for a sphere moving in the fluid, where the sphere's velocity is inversely proportional to the viscosity. However, below we will confirm this surprising property of chiral propulsion by an explicit computation.

One can also observe that the rotation is the simplest example of a ${\cal T}$-odd motion, but other movements of the body that break time reversal invariance could also lead to propulsion.
\vskip0.3cm

The key role of time-reversal invariance in propulsion of small chiral objects had been discussed in the seminal 1976 paper by Purcell \cite{purcell2014life}. To outline his argument, let us consider the Navier-Stokes equation describing the fluid motion around a moving object: 
\beq\label{NS}
- \nabla P + \eta \nabla^2 {\mathbf{v}} + \mathbf{f}_{ext} = \rho \left(\frac{\partial {\mathbf{v}}}{\partial t} + {\mathbf{v}}\ \nabla {\mathbf{v}} \right),
\eeq
where $P$ is the pressure, $\mathbf{f}_{ext}$ is an external force, $\mathbf{v}$ is the velocity of a fluid element, and $\eta$ and $\rho$ are the shear viscosity and the density of the fluid, respectively. Note that this equation contains an explicit dependence on time, and implies the breaking of time reversal invariance by viscous effects -- the r.h.s. of (\ref{NS}) is even under ${\cal T}$, while the term $\nabla^2 {\mathbf{v}}$ on the l.h.s. is ${\cal T}$ --odd (recall that velocity is $v=dx/dt$, so it changes sign under time reversal). This requires the shear viscosity $\eta$ to be odd under ${\cal T}$, consistent with our expectations for a quantity describing dissipation. Therefore the motion described by the Navier-Stokes equation in general breaks the time-reversal invariance. Shapere and Wilczek gave an elegant formulation of the ${\cal T}$-breaking character of swimmers at low Reynolds number in terms of an effective gauge field over the space of shapes \cite{shapere1989geometry}. 
\vskip0.3cm

Let us assume that a characteristic size of the object moving in the fluid is $L$. The motion of the object critically depends on the value of the Reynolds number
\beq
R = \frac{L {\rm{v}} \rho}{\eta} \equiv \frac{L \rm{v}}{\nu},
\eeq
where $\nu = \eta/\rho$ is the so-called kinematic viscosity. The Reynolds number represents the ratio of the inertial to the viscous forces, so that at small $R \ll 1$ the viscous forces dominate. 
Since the kinematic viscosity is given by the product $\nu = {\rm v}_T\ l$ of thermal velocity ${\rm v}_T$ and the mean free path $l$, the Reynolds number is the product of the ratio of the object velocity to the thermal velocity and the ratio of the object's size to the mean free path:
\beq
R = \frac{\rm{v}}{{\rm v}_T}\ \frac{L}{l}.
\eeq
The limit of small $R$ can be reached in any fluid of finite viscosity for sufficiently small and/or sufficiently slow objects, when $L\ \rm{v} \ll \nu$. 
\vskip0.3cm

For example, let us consider the motion of bacteria in water 
with kinematic viscosity $\nu \simeq 10^{-2}$ cm$^2$/s. The bacteria of a $\sim 1\ \mu m$ size (such as rod-shaped {\it Bellovibrio bacteriovorus}) can propel themselves by a flagellum 
that makes $\sim 100$ rotations per second, resulting in the velocity of about $\sim 100\ \mu m/s$. The corresponding Reynolds number is $R = L \rm{v}/\nu \sim 10^{-4}$. On the other hand, for a tuna fish of size $\sim 1\ m$ and velocity $\sim 10\ m/s$, the Reynolds number is very large, $R \sim 10^7$. 
\vskip0.3cm

The r.h.s. of Navier-Stokes equation (\ref{NS}) is of the order of $\rho \rm{v}^2/L$, whereas the second term on the l.h.s. is $\sim \eta\ v/L^2$; therefore their ratio is of the order of the Reynold number $R$.
This means that in the limit of small Reynolds number appropriate for slow motion and/or for small objects, the non-linear Navier-Stokes equation is reduced to a much simpler linear equation describing the Stokes flow, or "creeping flow":
\beq\label{Stokes}
- \nabla P + \nu \nabla^2 {\mathbf{v}} + \mathbf{f}_{ext} = 0 .
\eeq
This equation has no explicit dependence on time (other than through the time-dependent boundary conditions). The resulting flow can be found at any given moment of time without knowledge of the flow at any other times. Therefore any Stokes flow solution is also the solution of the time-reversed flow -- in other words, all solutions of (\ref{Stokes}) are invariant under time-reversal ${\cal T}$. This means that there is no entropy produced in the Stokes flow. A famous illustration of the reversibility of Stokes flow is the Taylor-Couette experiment in which the mixing of the ink in a viscous fluid by rotating concentric cylinders can be undone by reversing the direction of the mixer.
\vskip0.3cm

The time-reversal invariance of the Stokes flow implies that the propulsion of a rotating chiral object described by (\ref{propel}) can be performed in a non-dissipative way. This fact reveals a striking similarity of the classical problem of propulsion to the new class of macroscopic quantum phenomena induced by the so-called chiral and gravitational anomalies in quantum field theory, see \cite{kharzeev2014chiral} for review. In quantum field theory, the relation (\ref{propel}) arises from the rotation of a system with chiral fermions \cite{kharzeev2007charge,erdmenger2009fluid,landsteiner2011holographic,kharzeev2011testing}. The vector current (\ref{propel}) is generated if the densities of left- and right-handed fermions are different, i.e. when the parity invariance is broken ("the chiral vortical effect"). This breaking of parity can be achieved by coupling the fermions to an external gauge field configuration with a nontrivial topological contents characterized by a non-zero Chern-Simons number. In fluid dynamics, a non-zero chirality of the background can lead to the propulsion of chiral solitons along the vortices \cite{hirono2018dynamics}. We will now utilize the analogy to quantum physics to get a new insight into the problem of chiral propulsion at micro- and nano-scales.
 \vskip0.3cm
 
The rest of the paper is organized as follows. In Section \ref{sec:chiral_sort} we discuss the possible methods of chiral sorting (separation of enantiomers with opposite chirality) that utilize helical hydrodynamical flows and helical electromagnetic fields; some of them have been already realized experimentally. In Section \ref{sec:green_meth} we apply the Green's function method to evaluating the propulsion velocity at low Reynolds number, and illustrate it by deriving the Stokes formula for a spherical body propagating in a viscous fluid.  In Section \ref{sec:quant} we apply the Green's function method to the quantitative description of the experimental data on the separation of chiral nano-scale objects by electromagnetic fields. Finally, in Section \ref{sec:sum} we summarize our results. 

\section{Chiral sorting}\label{sec:chiral_sort}

Let us discuss the possible methods of separating the objects of different chirality in fluids. This is a problem of great importance -- for example, in pharmaceutics, since most of the new drugs are chiral, and it is necessary to obtain chirally pure drugs by separating different enantiomers. In biomedicine, propulsion of chiral drugs may enable precise drug delivery. 
\vskip0.3cm

Any method of chiral sorting has to rely on the breaking of parity invariance. We may divide the methods of chiral sorting in two broad groups: 
\vskip0.3cm

I) parity invariance ${\cal P}$ is broken only by the difference between the left and right enantiomers, whereas the flow and/or external electromagnetic fields are even under ${\cal P}$; 

II) parity invariance ${\cal P}$ is broken both by the enantiomers and by the flow and/or external electromagnetic fields.  
\vskip0.3cm

An example of type I is the chiral separation induced by propulsion (\ref{propel}): since the coefficient $\alpha$ in  (\ref{propel}) is odd under parity, the rotating left- and right-handed enantiomers will move in opposite directions. An ingenious way to make chiral objects rotate in a fluid was proposed by Baranova and Zel'dovich  \cite{baranova1978separation}, and utilizes rotating magnetic (or electric) fields. The interaction of rotating magnetic (electric) fields with the magnetic (electric) dipole moment of the chiral object induces on it a torque that makes the object rotate. Recently this theoretical proposal was realized experimentally using the colloid of nano-scale chiral screws \cite{schamel2013chiral}. In Section \ref{sec:quant} we will provide a quantitative theoretical description of the results of this experiment. 
\vskip0.3cm
 
Type II chiral sorting relies on a parity-odd hydrodynamical flow or parity-odd configuration of electromagnetic fields. In both cases, one can use the Chern-Simons three-form to describe the underlying topology; for hydrodynamical flows, it is called "kinetic helicity" and is given by
\beq\label{cs-kin}
{\cal{K}} \equiv \int d^3x\ {\mathbf{v}}\cdot {\mathbf{\Omega}},
\eeq
where ${\mathbf{v}}$ is the velocity of the fluid, and ${\mathbf{\Omega}} = \frac{1}{2}\ \nabla \times {\mathbf{v}}$ is vorticity. A well-known example of hydrodynamical flow characterized by a non-zero helicity (\ref{cs-kin}) is Beltrami flow (also known as Gromeka-Arnold-Beltrami-Childress flow) that is an exact solution of Euler's equation with  
${\mathbf{v}} \sim {\mathbf{\Omega}}$. On symmetry grounds one can expect that left- and right-handed enantiomers will interact differently with the helical Beltrami flow. A recent experiment 
\cite{aristov2013separation} indeed observed the separation of chiral colloidal particles in a helical hydrodynamical flow. 
\vskip0.3cm

Another possible kind of type II sorting (that has been explored to a much smaller extent) uses helical external electromagnetic fields characterized by "magnetic helicity" 
\beq\label{cs-mag}
{\cal{H}}_M \equiv \int d^3x\ {\mathbf{A}}\cdot {\mathbf{B}},
\eeq 
and its dual
\beq\label{cs-el}
{\cal{H}}_E \equiv \int d^3x\ {\mathbf{C}}\cdot {\mathbf{E}} ,
\eeq
where ${\mathbf{A}}$ is the vector gauge potential, ${\mathbf{B}} = \nabla \times {\mathbf{A}}$ is the magnetic field, ${\mathbf{E}}$ is the electric field, and the dual gauge potential ${\mathbf{C}}$ is defined by ${\mathbf{E}} = \nabla \times {\mathbf{C}}$. An alternative, and explicitly gauge invariant, measure of chirality can be obtained by using the electric and magnetic fields instead of the gauge potential; the resulting "optical chirality" \cite{tang2010optical, tang2011enhanced} is given by
\beq
C \equiv \int d^3 x\ \left[ \frac{\epsilon}{2}\ {\mathbf{E}}\cdot \nabla \times {\mathbf{E}} + \frac{1}{2 \mu}\ {\mathbf{B}}\cdot \nabla \times {\mathbf{B}} \right],
\eeq
where $\epsilon$ and $\mu$ are the permittivity and permeability, respectively.
\vskip0.3cm

Let us mention some possible realizations of helical electromagnetic fields. Perhaps the simplest is the circularly polarized light. Recently the circularly polarized laser radiation has been utilized for optofluidic chiral sorting \cite{tkachenko2014optofluidic}. 
\vskip0.3cm

Another possibility to realize the optofluidic chiral sorting is to use the parallel electric magnetic and fields. Note that 
\beq
\int d^3x\ {\mathbf{E}}\cdot {\mathbf{B}} = - \frac{1}{2} \frac{\partial {\cal{H}}_M}{\partial t},
\eeq
so that static parallel electric and magnetic fields pump chirality into the system. The sign of chirality imbalance can be reversed by switching to the anti-parallel fields. 
One can also combine a static magnetic or electric field with a rotating one, creating non-zero ${\cal{H}}_M$ or ${\cal{H}}_E$. The field configuration with parallel electric and magnetic field can be also engineered optically by combining two counter-propagating laser beams with the opposite circular polarizations \cite{bliokh2014magnetoelectric}.
\vskip0.3cm

In any case, to evaluate the resulting propulsion velocities we need a method of solving Stokes equation for a chiral body under the influence of an external force. In the next Section we will utilize for this purpose the Green's function approach that is enabled by the linearity  of the Stokes equation (\ref{Stokes}).

\section{Propulsion velocity: the Green's function method}\label{sec:green_meth}

\subsection{Green's function method for the Stokes equation}

Let us describe the flow of fluid around a rigid body by using the Stokes equation (\ref{Stokes}) supplemented by the fluid incompressibility condition
\begin{equation}
	\label{eq2}
	\nabla\cdot \mathbf{v}(\mathbf{r})=0 .
\end{equation}
The linearity of the Stokes equation makes it convenient to solve it using the Green's function method. To apply it to (\ref{Stokes}), one considers the response of the fluid to an external point force $\mathbf F$ acting at the origin $\mathbf{r}=0$, 
\beq
- \nabla P + \nu \nabla^2 {\mathbf{v}} = - \mathbf{F}\ \delta(\mathbf{r}),
\eeq
and evaluates the corresponding Green's function, called the Oseen tensor, 
\begin{equation}
	\label{eq3}
	{G}_{ij}(\mathbf{r})=\frac{ {\delta }_{ij} }{ r }+\frac{ {r}_{i}{r}_{j} }{ {r}^{3} };
\end{equation}
assuming that both the pressure and the fluid flow vanish at infinity; 
for recent applications of this method see \cite{manghi2006hydrodynamic,smith2017nearest,marin2012boundary,lisicki2013four}. 
For a point force, the corresponding solution for the velocity field 
\beq
{v}_{i}(\mathbf{r}) = \frac{1}{ 8\pi \nu }\ F_j\ {G}_{ij}(\mathbf{r})
\eeq
is called Stokeslet; the corresponding solution for pressure is 
\beq
P(\mathbf{r}) = \frac{\mathbf{F}\ \mathbf{r}}{4 \pi r^3}.
\eeq
Consider now a rigid body immersed in a fluid. Let us assume that it exerts on the fluid a continuous force distribution with density $\mathbf{f}(\mathbf{r})$ localized on the surface of the body; the fluid velocity is then given by
\begin{equation}
	\label{eq4}
	{v}_{i}(\mathbf{r})=\frac{1}{ 8\pi \nu }\int_{S}{ {G}_{ij}(\mathbf{r}-\mathbf{r}\,'){f}_{j}({r}')d{\sigma }' }
\end{equation}
where the integration is over the surface $S$ of the body.
\vskip0.3cm

Let us impose the usual "no-slip" boundary condition on the surface of the body. In this case 
the velocity of the body is equal to the velocity of the fluid on its surface:
\begin{equation}
	\label{eq5}
	\mathbf{v}_{body}=\lim_{ \mathbf{r}\to \mathbf{r}\,' } \mathbf{v}(\mathbf{r}),
\end{equation}
where ${\mathbf{r}\,' }$ can be chosen as any point on the body's surface.
This trivial observation, 
combined with the Green's function method outlined above, allows 
to simplify the computation of the propulsion velocity, as will now show.

\subsection{Green's function method for a sphere in a viscous fluid}\label{sphere}

To illustrate the method that we will later use to compute the propulsion velocity, let us first consider the classic problem solved by Stokes in 1851 and find the velocity of a sphere moving in a fluid at small Reynolds number. Let us assume, following Stokes, that there is an external force $\mathbf{F}^{ext}$ applied to the sphere of radius $R$ along the axis $z$. Since we are interested in a uniform motion, the first Newton's law tells us that this external force is balanced by the uniformly (due to the symmetry of the problem, see \cite{falcovich2011fluid}) distributed drag force with density $f_z$ exerted on the sphere by the viscous fluid:
\begin{equation}
	\label{eq6}
	{f}_{z}=- \frac{F_{z}^{ext}}{S},
\end{equation}
where $S=4\pi R^2$ is the surface area; see Fig. \ref{sphere}.
\begin{figure}[h]
	\begin{center}
		\includegraphics[scale=0.6]{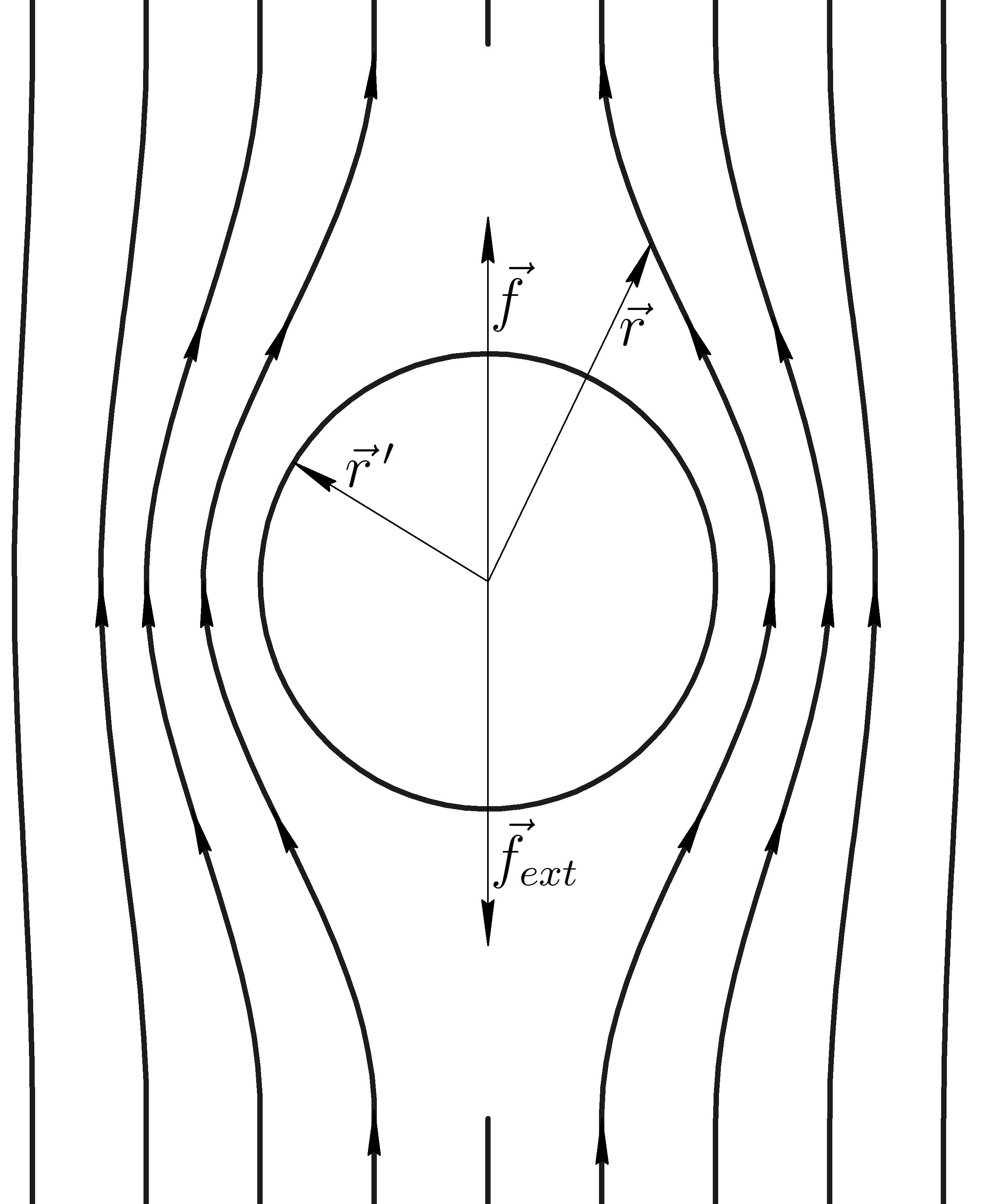}
		\caption{The viscous fluid flow around the sphere.}
		\label{sphere}
	\end{center}
\end{figure}

The method that we propose is the following: i) use the Green's function (Stokeslet) to write down the expression for the velocity of the fluid, and then ii) evaluate this velocity at the surface of the object, at a point where the symmetry of the object makes this evaluation the easiest.

To implement this, let us first write down the component of the velocity 
 field along the axis $z$ using (\ref{eq4}):  
\begin{equation}
	\label{eq8}
	v_{z}(\mathbf{r})=\frac{1}{ 8\pi \eta }\int_{S}{\left[\frac{\delta_{zj}}{|\mathbf{r}-\mathbf{r}'|} + \frac{(r_{z}-r'_{z})(r_{j}-r'_{j})} {|\mathbf{r}-\mathbf{r}'|^3} \right] f_{j}(r')d\sigma'}
\end{equation}
The symmetry of the problem makes it easiest to evaluate the integrals in (\ref{eq8}) at points located along the axis of the sphere; 
for positive $z>R$, after performing simple integrations, we find
\begin{equation}
	\label{eq13}
	v_{z}= - \frac{1}{8\pi \eta }\frac{F_{z}}{4\pi R^{2}} \left( \frac{4\pi R^2}{z} + \frac{4 \pi R^2 z}{z^2 - R^2} + 2 \frac{4 \pi R^4}{z (R^2-z^2)} - \frac{4 \pi R^4 (2 R^2 + z^2) }{3 z^3 (R^2-z^2)} \right)= \frac{F_{z} (R^2 - 3z^2)}{12 \pi \eta z^3}
\end{equation}
Now, in order to find the velocity of the sphere, according to Eq.~(\ref{eq5}) we evaluate the velocity of the fluid at the upper tip of the sphere: 
\begin{equation}
	\label{eq14}
	{v}^{sphere}_{z}=\lim_{z \to R} v_{z}=\lim_{z \to R} \frac{F_{z} (R^2 - 3z^2)}{12 \pi \eta z^3}= -\frac{F_{z}}{6 \pi \eta R} =  \frac{F_{z}^{ext}}{6 \pi \eta R}
\end{equation}
This is the Stokes' law for the velocity of a sphere moving in a viscous fluid at low Reynolds number.

\section{Chiral propulsion at nano-scales by electromagnetic fields}\label{sec:quant}

 \subsection{Hydrodynamics of helical bodies}\label{sec:hel}
 
 Let us now proceed to the hydrodynamics of rigid helical bodies driven by an external torque. Recently such experiments have been carried out with nano-scale screws driven by rotating electric fields \cite{schamel2014nanopropellers, schamel2013chiral}, motivated by the theoretical proposal  \cite{baranova1978separation}. The propulsion of helical bodies is a problem of great interest in biology, as this is the underlying mechanism of locomotion for bacteria and spermatozoa \cite{cox1970motion,lauga2009hydrodynamics,he2014propulsive, purcell1997efficiency,marin2012boundary}, see \cite{elgeti2015physics} for a review. This problem is also of direct relevance for the separation of chiral enantiomers in pharmacology. 
 
  \begin{figure}[h]
 	\begin{center}
 		\includegraphics[scale=0.6]{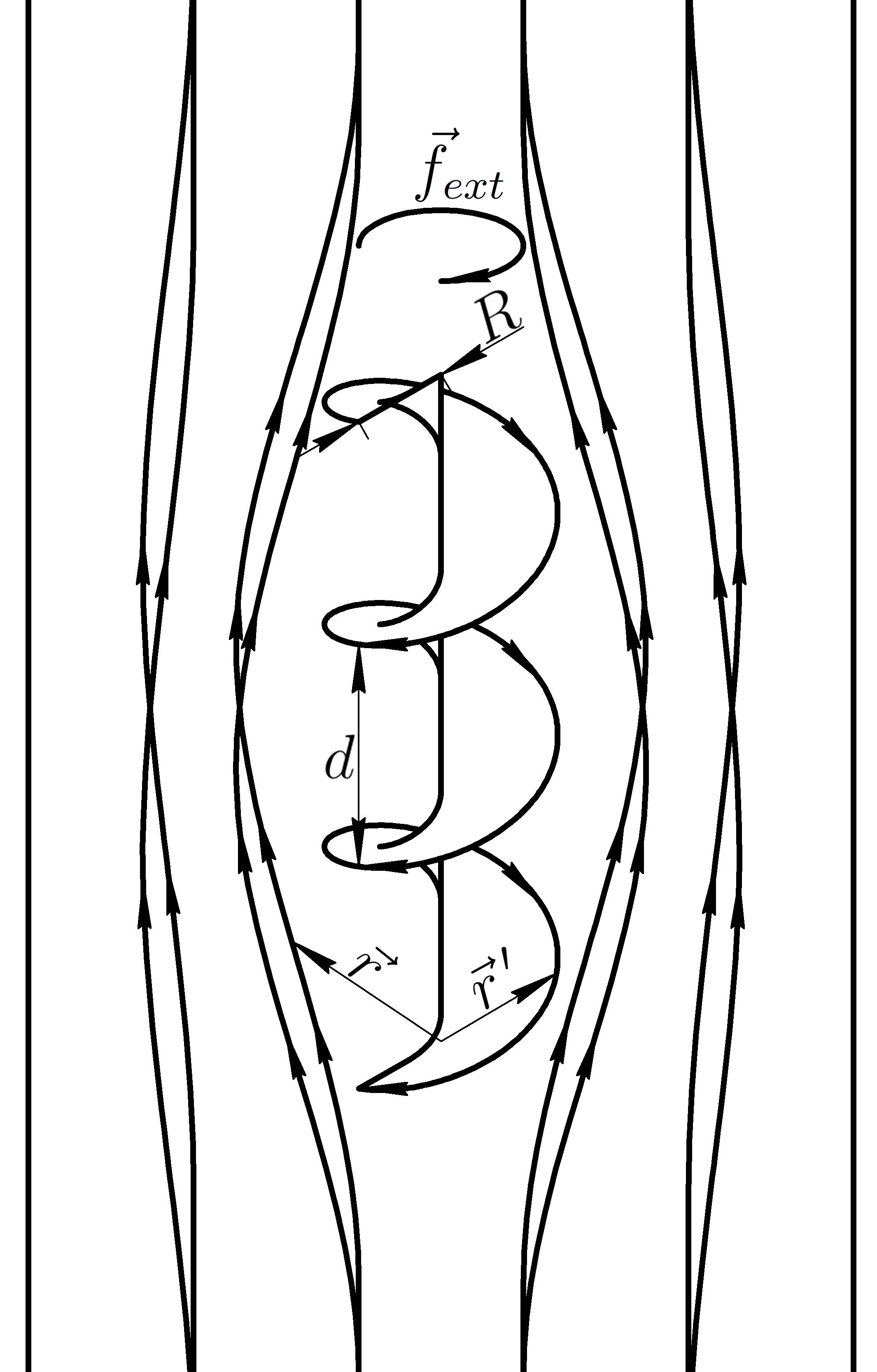}
 		\caption{The propulsion of a rotating helicoid in the fluid.}
 		\label{helicoid}
 	\end{center}
 \end{figure}
 
 Consider a helix with a torque applied to it along the symmetry axis that makes it rotate with the angular frequency $\omega$. Due to the breaking of parity, we then expect, according to (\ref{propel}) that the helix will start moving along the axis of the angular momentum, which is also the symmetry axis, with a velocity ${\rm v} \sim \omega$. Our task is to evaluate the coefficient in this relation, assuming that the Reynolds number is small and the Stokes equation applies.
 \vskip0.3cm
 
 We will again use the Green's function method outlined above; however, it appears that there is an important difference between the treatment of parity-symmetric and parity-odd rigid bodies in this approach. Namely, for the case of a symmetric body, such as a sphere considered in section \ref{sphere}, we have not suffered from the short-distance singularities in the evaluation of the integrals involving the Green's functions, even though the Green's function itself (the Stokeslet) is singular. The cancelation of the singular terms can be traced back to the symmetry of the body. On the other hand, for helical bodies, this symmetry is not present, and the singularities do not cancel. 
 \vskip0.3cm
 
 The situation here is quite similar to what happens in relativistic quantum field theory -- the interactions of external gauge fields with chiral fermions present a short-distance divergence that cannot be regularized in a way consistent with a global chiral symmetry; this leads to the so-called chiral anomaly 
 \cite{adler1969axial,bell1969pcac}. Hydrodynamics is an effective theory, and the application of the Green's function method in this case requires an explicit regularization. Such a regularization in hydrodynamics has been proposed by Cortez \cite{cortez2005method}, who has replaced the pointlike source by the smeared distribution \begin{equation}
 \phi_\epsilon(\mathbf{r}-\mathbf{r} ')=\frac{15\epsilon^4}
 {8\pi (\rho^2+\epsilon^2)^\frac{7}{2}},
 \end{equation}
 where $\rho=\|\mathbf{r}-\mathbf{r}'\|$,
with the corresponding regularized Oseen tensor
 \begin{equation}
 S_{ij}^{\epsilon}(\mathbf{r},\mathbf{r}')=
 \delta_{ij}\frac{\rho^2+2\epsilon^2}{(\rho^2+\epsilon^2)^\frac{3}{2}}+
 \frac{(r_i-r_i')(r_j-r_j')}{(\rho^2+\epsilon^2)^\frac{3}{2}}.
 \end{equation}
 The fluid velocity field is then given by 
 \begin{equation}
 \int v_i(\mathbf{r}')\phi_\epsilon(\mathbf{r}-\mathbf{r}')dV'=
 \frac{1}{8\pi \eta}\int S_{ij}^{\epsilon}(\mathbf{r},\mathbf{r}')f_j(\mathbf{r}')dS'.
 \end{equation}
 For distances to the surface of the body that are larger than $R_c=\sqrt{\frac{5\epsilon}{2}}$, 
 the regularization error is $O(\epsilon^2)$; at shorter distances $0\leq dist(\mathbf{r},\Sigma) < R_c$, the error is $O(\epsilon)$. We are interested in evaluating the fluid velocity at the surface of the body, where the l.h.s. integral yields
 \begin{equation}
 \int v_i(\mathbf{r}')\phi_\epsilon(\mathbf{r}-\mathbf{r}')dV'=v_i(\mathbf{r})+O(\epsilon).
 \end{equation}
 \vskip0.3cm
 
 Let us now specify the shape of the body. As a prototype example of a helical body we will consider a  a finite helicoid, parametrized by
 \begin{equation}
 \begin{cases}
 \begin{matrix}
 x=r\cos\theta ;     & 0\leq r\leq R\\
 y=r\sin \theta ;           &0\leq \theta \leq 2\pi\\
 z=\frac{d\theta}{2\pi},
 \end{matrix}
 \end{cases}
 \end{equation}
 where $R$ is the radius of helicoid and $ d $ is its pitch.
 \vskip0.3cm
 
 One more thing to specify is the force density distribution around the surface. We have already assumed that our rigid body rotates with the angular velocity $\omega$, and we will assume that 
 this angular velocity is created by a uniformly distributed force. Let us choose $z$ as the symmetry axis of the helicoid (see Fig. \ref{helicoid}); we do not push the helicoid along $z$, so that the applied force has only $x$ and $y$ components:
  \begin{equation}
 \mathbf{f}=\frac{F}{S}{\{\sin\theta, \cos\theta , 0 \}}.
 \end{equation} 
 Knowing this external force and the parametric equation for the surface we can now solve the Stokes equation and find the fluid velocity using the regularized Stokeslet:  
 \begin{equation} \label{eq: v general}
 v_i(\mathbf{r})=
 \frac{1}{8\pi\eta}\iint S_{ij}^\epsilon(\mathbf{r}-\mathbf{r}')f_j(r') d\sigma',
 \end{equation}
 where the regularization error is of order $\epsilon$. The velocity of the elements of the helicoid's surface are fixed by 
 \begin{equation*}
 \bf{v}=[\mathbf{\omega}\times \mathbf{r}],
 \end{equation*}
 where $\mathbf{w}$ is the angular velocity and $\mathbf{r}$ is the distance from the symmetry axis;
 for example, along the x-axis
 \begin{center}
 	$v_y(x)=\rm{\omega}\ x $.
 \end{center}
 On the other hand, for the fluid velocity we have
 \begin{equation*} \begin{split}
 v_y=\frac{1}{8\pi \eta}\iint S_{yj}^\epsilon(\mathbf{r}-\mathbf{r}')f_j(\mathbf{r}')d\sigma'=\nonumber\\
 =\frac{1}{8\pi\eta}\iint\bigg( \frac{(\rho^2+2\epsilon^2)f_y}{(\rho^2+\epsilon^2)^{\frac{3}{2}}}+\frac{(y-y')(r_j-r_j')f_j}{(\rho^2+\epsilon^2)^{\frac{3}{2}}}\bigg)d\sigma' .
 \end{split}
 \end{equation*}
 Along the $x$ axis, $\mathbf{r}=(x,0,0)$,
 \begin{equation} \label{vy1}
 \begin{split}
 v_y(x)=\frac{1}{8\pi\eta}\iint \bigg(\frac{f_y(\mathbf{r}')(\rho^2+2\epsilon^2)}{\rho^2+\epsilon^2}+\\
 +\frac{(x'y'-xy')f_x(\mathbf{r}')}{(\rho^2+\epsilon^2)^{\frac{3}{2}}}+
 \frac{y'y'f_y(\mathbf{r}')}{(\rho^2+\epsilon^2)^{\frac{3}{2}}}\bigg)d\sigma' ,
 \end{split}
 \end{equation}
 where the element of the surface is 
 \begin{equation}
 d\sigma=\sqrt{\Big(\frac{d}{2\pi}\Big)^2+r^2}drd\theta.
 \end{equation}
 Setting the point $\mathbf{r}$ on the x-axis we can express 
 \begin{equation*}
 \rho=\|\mathbf{r}-\mathbf{r}'\|=\sqrt{r'^2+x^2+\left(\frac{d \theta'}{2\pi}\right)^2 - 2xr'\cos \theta'}.
 \end{equation*}
 Since we assume no-slip boundary conditions, the velocity of fluid at the surface is equal to that of the helicod; we thus get a matching condition
 \begin{equation}\label{match}
 \lim_{x\rightarrow R}v_y(x) =\omega R.
 \end{equation}
 We use this condition to express the uniformly distributed force in terms of the angular frequency ${\rm \omega}$, for a given geometry of the helicoid. 
 \vskip0.3cm
 
 Let us now proceed to evaluating the propulsion velocity along the $z$ axis; it is given by an analogous expression
 \begin{equation} \label{eq: v_z}
 v_z(z)=\frac{1}{8\pi \eta }\int \int \frac{(z-z')(r_j-r_j')f_j(\mathbf{r})'}{(\rho^2+\epsilon^2)^{\frac{3}{2}}}d\sigma' .
 \end{equation}
 Let us choose a point on the $z$ axis, with $\mathbf{r}=(0,0,z)$; we get
 \begin{equation}\label{prop_vel}
 v_z(z)=\frac{1}{8\pi\eta}\int \int \frac{(z'x'-zx')f_x(\mathbf{r}')+(z'y'-zy')f_y(\mathbf{r}')}{(\rho^2+\epsilon^2)^{\frac{3}{2}}}d\sigma' ,
 \end{equation}
 with
 \begin{equation*}
 \rho=\sqrt{r'^2+z^2+\Big( \frac{d\theta}{4\pi}\Big)^2-\frac{dz\theta}{\pi}} .
 \end{equation*}
 Again, if we approach the surface of helicoid, the fluid velocity of fluid has to be equal to that of the helicoid due to the no-slip boundary condition. The unknown force can be expressed from the equation (\ref{vy1}); as a result, the propulsion velocity of the helicoid is completely determined: along z-axis as a function of angular velocity and geometry. 
 \begin{equation} \label{omegaY}
 v_z= \alpha \ {\rm{\omega}}, 
 \end{equation}
 where the constant $\alpha$ depends on the geometry of helicoid and can be found by performing  integrals in (\ref{prop_vel}) and (\ref{vy1}). 
 The explicit expression for $\alpha$ is cumbersome, and we will present below the numerical results.
 \vskip0.3cm
 
 An interesting feature of our result is the following: because the force density $f$ found from the matching condition appears proportional to viscosity $\eta$, upon the substitution of $f$ in the expression for the propulsion velocity (\ref{prop_vel}) we find that the dependence on the value of viscosity cancels out. Therefore, the propulsion velocity is entirely determined by the angular velocity of rotation and the geometry of the body, and {\it does not depend on the viscosity of the fluid} as long as the Reynolds number remains small. This is in sharp contrast with the Stokes law ({eq14}) for a (non-chiral) sphere, in which the velocity of motion is inversely proportional to the fluid's viscosity.
 \vskip0.3cm
 
 Note however that the independence of the propulsion velocity on viscosity (at low Reynolds number) is in accord with our analysis of the properties of (\ref{propel}) under the time-reversal ${\cal T}$: since $\alpha$ has to be even under time reversal, it cannot be inversely proportional to the  ${\cal T}$-odd viscosity $\eta$. This difference between the Stokes law and the propulsion velocity (\ref{propel}) originates entirely from the chirality.
  
 \subsection{Helicoid with a sphere}
 
 The chiral colloids used in the experiment  \cite{schamel2013chiral} may be reasonably well approximated by a helicoid with a sphere attached to it; see the image in Fig.2 of  \cite{schamel2013chiral}. We will thus model the chiral colloids of  \cite{schamel2013chiral} by rigidly attaching a sphere of radius  $R_s$ to the helicoid considered. 
 To perform the computation, we could simply add the sphere in the surface integrals of section \ref{sec:hel}. However the computation can be simplified by noting that rotation of the sphere does not lead to propulsion, and the only effect of the sphere is to add to the viscous drag force in the $z$ direction. We can thus add the resulting drag force to the force density distribution:
 \begin{equation}\label{force}
 \mathbf{f}=\frac{1}{S}{\{F\sin\theta;F\cos\theta; -6\pi R_s \eta v_z \}}.
 \end{equation} 
 Since $f_z$ is no longer zero, the formulas for the components of velocity get modified:
 \begin{equation} \label{v_y sphere}
 \begin{split}
 v_y(x)=\frac{1}{8\pi\eta}\ \iint S_{yj}^\epsilon(\mathbf{r}-\mathbf{r}')f_j(\mathbf{r}')d\sigma'=\Arrowvert \mathbf{r}=(x,0,0) \Arrowvert=  \\
 =\iint  \frac{(\rho^2+2\epsilon^2)f_y(\mathbf{r}')}{(\rho^2+\epsilon^2)^{\frac{3}{2}}}d\sigma'+\iint \frac{(y'x'-xy')f_x+y'y'f_y+y'z'f_z}
 {(\rho^2+\epsilon^2)^{\frac{3}{2}}}d\sigma'.
 \end{split}
 \end{equation}
 \begin{equation} \label{v_z sphere}
 \begin{split}
 v_z(z)=\frac{1}{8\pi\eta}\ \iint  G_{zj}(\mathbf{r}-\mathbf{r}')f_j(\mathbf{r}')d\sigma'=\Arrowvert \mathbf{r}=(0,0,z) \Arrowvert= \\
 =\iint  \frac{(\rho^2+2\epsilon^2)f_z}{(\rho^2+\epsilon^2)^{\frac{3}{2}}}d\sigma'+\\
 +\iint \frac{(z'x'-zx')f_x+(z'y'-zy')f_y+(zz-2zz'+z'z')f_z}
 {(\rho^2+\epsilon^2)^{\frac{3}{2}}}d\sigma'  .
 \end{split}
 \end{equation}
 As before, we find the force $F$ in (\ref{force}) matching the velocity given by the equation (\ref{v_y sphere}) to the velocity of the rigid body by Eq. (\ref{match}). We then substitute the value of $F$ into 
(\ref{v_z sphere}) and find the relation similar to (\ref{omegaY}), but with a different coefficient $\alpha'$ determined by the modified geometry:
 \begin{equation} 
 v_z =  \alpha'\ \omega .
 \end{equation}
 
 \subsection{Numerical results}
 
 In this section we present numerical results for the problems described above. We have computed the propulsion velocities for the helicoid rotating around its long axis, see Table(\ref{table:1}), and for the rotating helicoid with a rigidly attached sphere, see Table(\ref{table:2}). As described above, the computations have been done by using the method of regularized Stokeslet, with the short-distance cutoff for the hydrodynamical description of the order of the mean free path of water molecules, i.e. $\epsilon = 10^{-10}\ m$. Note however that when we express the force from the expression for the in-plane velocity and substitute it in the expression for the propulsion velocity, the singular dependence on the cutoff $\epsilon$ cancels out. As we will discuss below, this procedure is similar to renormalization in relativistic quantum field theory.

 \begin{table}[h!]
 	\centering
 	\begin{tabular}{|c| c| c| c|} 
 		\hline
 		$d ,\mu m$ & $R, \mu m$ & $v_z, \frac{\mu m}{s} (\theta_0=2\pi)$ & $v_z, \frac{\mu m}{s} (\theta_0=6\pi)$ \\ [0.5ex] 
 		\hline
 		0.7 & 0.25 & 1.29 & 1.35 \\ 
 		0.7 & 0.3 & 1.33 & 1.44 \\
 		0.7&0.35&1.3&1.49\\ \hline
 		0.6 & 0.25 & 1.14 & 1.23 \\
 		0.6 & 0.3 & 1.11 & 1.27 \\
 		0.6 & 0.35&1.03&1.28\\ \hline
 		0.5&0.25&0.93&1.06\\
 		0.5&0.3&0.84&1.07\\
 		0.5 & 0.35 & 0.71 & 1.04\\  [0.5ex] 
 		\hline
 	\end{tabular}
 	\caption{The propulsion velocity $v_z$ of the helicoid for different geometries at the rotation frequency of $\omega=$20\ Hz. $R$ is the radius of the helicoid, and $d$ is the pitch. The column marked $\theta_0=2\pi$ corresponds to a single-period helicoid of length equal to the pitch, and the column $\theta_0=6\pi$ - the triple-period helicoid of length $3d$. }
 	\label{table:1}
 \end{table}
 
 \begin{table}[h!]
 	\centering
 	\begin{tabular}{|c| c| c| c|} 
 		\hline
 		$d ,\mu m$ & $R, \mu m$ & $v_z, \frac{\mu m}{s} (\theta_0=2\pi)$ & $v_z, \frac{\mu m}{s} (\theta_0=6\pi)$ \\ [0.5ex] 
 		\hline
 		0.7 & 0.25 & 0.69 & 0.94 \\ 
 		0.7 & 0.3 & 0.75 & 1\\
 		0.7&0.35&0.76&1.1\\ \hline
 		0.6 & 0.25 & 0.59 & 0.23 \\
 		0.6 & 0.3 & 0.61 & 0.88 \\
 		0.6 & 0.35&0.59&0.9\\ \hline
 		0.5&0.25&0.47&0.69\\
 		0.5&0.3&0.45&0.71\\
 		0.5 & 0.35 & 0.4 & 0.71\\  [0.5ex] 
 		\hline 
 	\end{tabular}
 	\caption{Same as Table \ref{table:1}, but for the helicoid with a sphere of radius $R_s=0.2 \mu$m attached to it.}
 	\label{table:2}
 \end{table}

\begin{figure}[h]
	\begin{center}
		\includegraphics[scale=0.8]{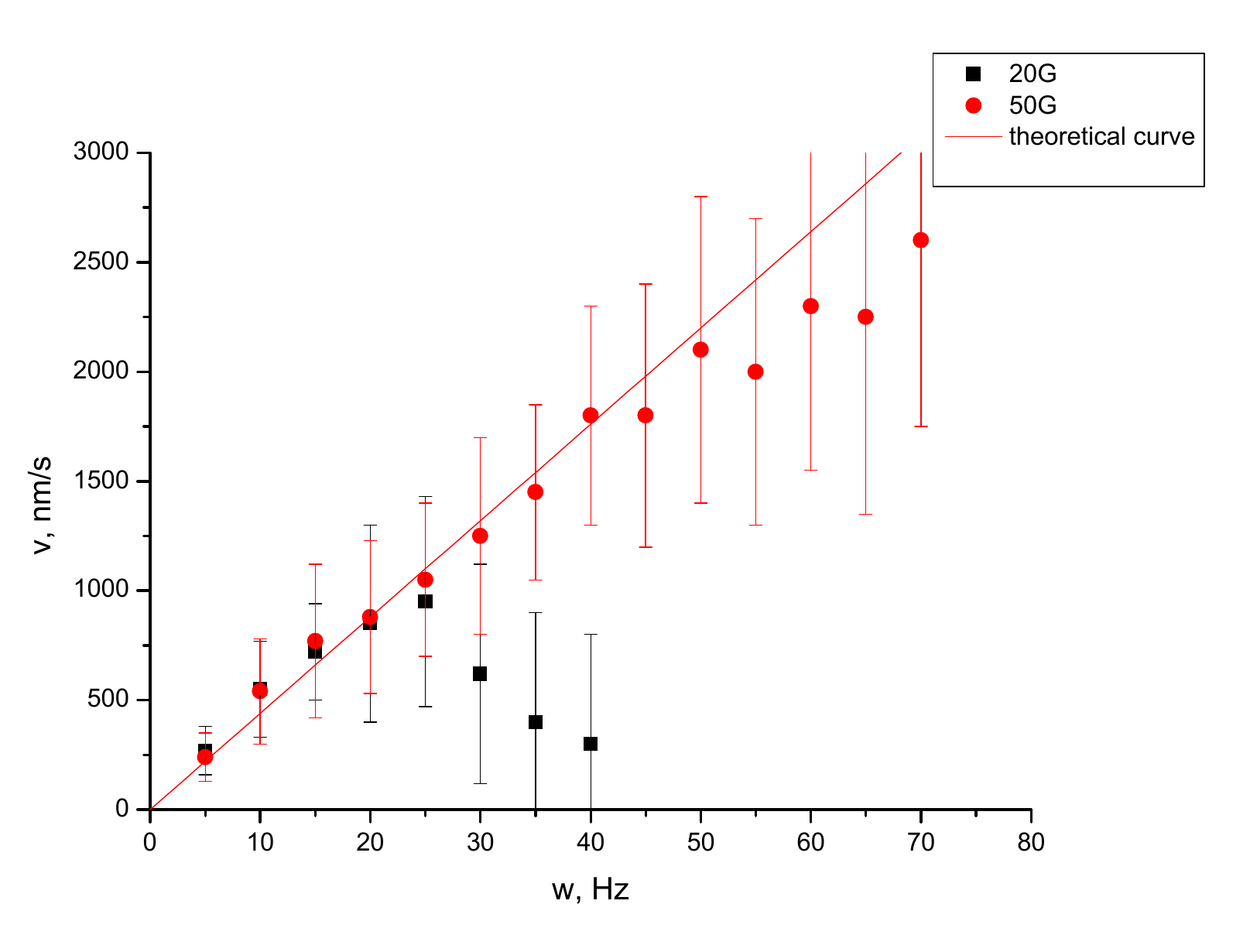}
		\caption{Comparison of our results for the propulsion velocity to the experimental data of Ref. \cite{schamel2013chiral} for magnetic fields of $B = 20$ G (black boxes) and $B = 50$ G (red circles).}
		\label{comparison}
	\end{center}
\end{figure}
 
Our model of the helical body with a sphere was designed to describe the colloidal chiral objects 
studied in a recent experiment by Schamel et al \cite{schamel2013chiral}. In this work, the chiral dipolar colloidal objects were brought in rotation by magnetic field rotating with frequency $\Omega = 5 - 80$ Hz, and the chiral separation was clearly observed. In Fig. \ref{comparison}, we compare our theoretical results for the helicoid with an attached sphere, of sizes chosen to model the shape of the colloidal objects used in the experiment, to the data  \cite{schamel2013chiral}. We have assumed that the body rotates with the same frequency as external magnetic field $\omega=\Omega$. This assumption is valid  below the "step-out" rotation frequency \cite{schamel2013chiral} 
 \begin{equation}\label{stepout}
 \Omega_{\rm step-out} \equiv \frac{M_{rem} B}{X \eta},
 \end{equation}
 determined by the value of magnetic field $B$,  the remnant magnetization $M_{rem}$ of the dipolar colloidal object,  viscosity $\eta$, and a dimensionless geometry factor $X$. At higher frequencies, the helicoid cannot rotate with the angular frequency of magnetic field, that is seen to occur for the weaker magnetic field of $B=20$ G, see Fig. \ref{comparison}.
\vskip0.3cm

We have found the optimal ratio of the pitch to the radius of the helicoid that maximizes the propulsion velocity; for a single-period helicoid, this optimal ratio is $R \simeq 0.43\ d$. We can define the absolute maximum velocity of the helicoid as $v_{max}=\frac{d\omega}{2\pi}$ -- this is the velocity with which the screw with pitch $d$ rotating with a frequency $\omega$ would move in a solid body. In units of this  $v_{max}$, the velocity $v_z$ achievable for the single-period helicoid described above is  $v_z/v_{max} \simeq 0.59$, so the propulsion is quite efficient. 

\begin{figure}[h]
 	\begin{center}
 		\includegraphics[scale=0.7]{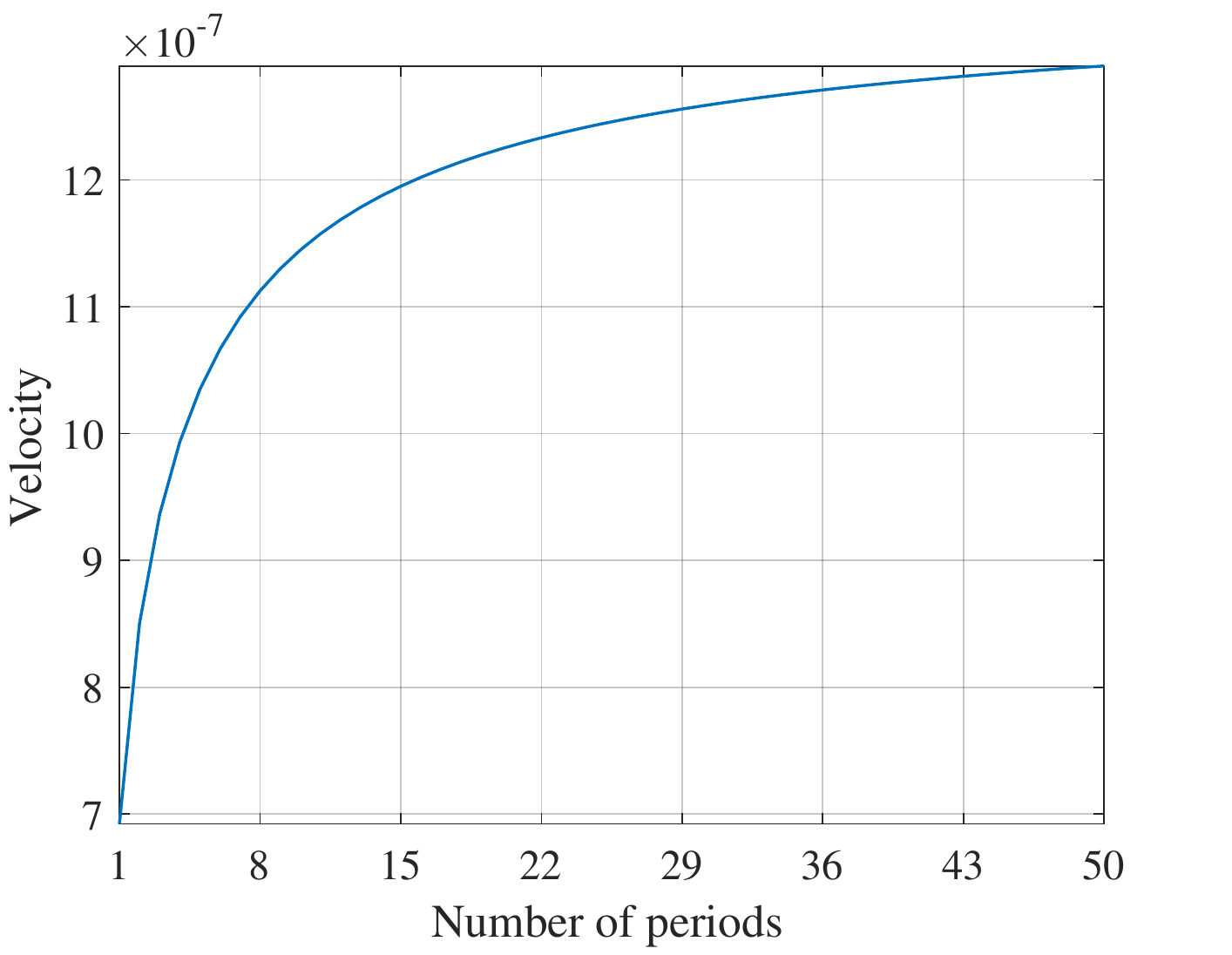}
 		\caption{The dependence of the propulsion velocity (in units of 100 nm/s) on the number of periods of the helicoid with a sphere, with rotation frequency of $\omega=$20\ Hz. The pitch of the helicoid is $d=0.7\ \mu$m, its radius is $R=0.25\ \mu$m, and the radius of the attached sphere is $R_s=0.2\ \mu$m.}
 		\label{thetadependence}
 	\end{center}
 \end{figure}

For the triple-period helicoid with the attached sphere of radius $R_s = 0.2\ \mu$m and the pitch of $d=0.7\ \mu$m, the optimal radius is $R = 0.55\ d$; the corresponding velocity is $v_z/v_{max} \simeq 0.54$ -- this means that the attached sphere does not significantly slow down the propulsion. 
It is also interesting to note that the propulsion velocity increases with the number of periods of the helicoid, but saturates and becomes constant for long helicoids, see Fig. \ref{thetadependence}.

\section{Summary}\label{sec:sum}

Our study of chiral propulsion at low Reynolds number has revealed some interesting properties of this phenomenon. The most intriguing of them is the independence of the propulsion velocity of helical bodies on the viscosity of the fluid, as long as the Reynolds number remains small. This phenomenon appears surprisingly similar to the chiral anomaly \cite{adler1969axial,bell1969pcac} in relativistic quantum field theory: the short-distance divergence yields a physical relation between the rate of generation of chiral charge and the rate of change of Chern-Simons number of the gauge field; this relation does not depend on the short-distance cutoff. In the framework of hydrodynamics, the short-distance cutoff is determined by viscosity. We have found that the computation of the velocity field (analog of the gauge field) suffers from the short-distance divergence, but the dependence on the short-distance cutoff disappears when we evaluate the propulsion velocity. As a result, the propulsion velocity appears independent of the viscosity. Note that the distribution of velocity field for the chiral propulsion is characterized by the topological Chern-Simons number that is generated with the rate proportional to the rotation velocity of the helicoid -- so the analogy to the chiral anomaly, and to the associated with it "chiral vortical effect"  \cite{kharzeev2007charge,erdmenger2009fluid,landsteiner2011holographic,kharzeev2011testing} is indeed quite close.
\vskip0.3cm

The independence of the propulsion velocity on viscosity (at low Reynolds number) also agrees with our analysis of the properties of (\ref{propel}) under the time-reversal ${\cal T}$. Since the coefficient $\alpha$ that determines the propulsion velocity has to be even under time reversal, it cannot be inversely proportional to the  ${\cal T}$-odd viscosity $\eta$. This striking difference between the Stokes law and the propulsion velocity (\ref{propel}) originates from chirality.
\vskip0.3cm

We applied the Green's function method to the description of experimental results \cite{schamel2013chiral} on the propulsion of chiral colloids by rotating magnetic fields, and found a good agreement with the data. Our analysis of the efficiency of chiral propulsion for different geometries of the helicoid may be useful for designing microfluidic devices. In particular, the model of the helicoid with an attached sphere may be used for designing drug delivery microdevices. The independence of the propulsion velocity on viscosity also suggests a way to increase the efficiency of chiral sorting by electromagnetic fields: since the maximal step-out rotation frequency (\ref{stepout}) is inversely proportional to viscosity, to maximize the propulsion velocity one should use the fluids with the smallest available viscosity for which the Reynolds number is still small. 

\vskip0.3cm

This work was initiated during the 2017 International competition in physics for students in Moscow that was supported by Skoltech. We thank G. Falkovich, L. Levitov, and \mbox{A. Molochkov} for very useful discussions. The work of D.K. was supported in part by the US Department of Energy under contracts No. DE-FG-88ER40388, DE-SC-0017662 and DE-AC02-98CH10886.
\vskip0.5cm

\bibliography{References}
\end{document}